# Secure and Distributed Assessment of Privacy-Preserving GWAS Releases[*]

Túlio Pascoal[†]
University of Luxembourg
Luxembourg
tulio.pascoal@uni.lu

Jérémie Decouchant[†]
Delft University of Technology
The Netherlands
j.decouchant@tudelft.nl

Marcus Völp
University of Luxembourg
Luxembourg
marcus.voelp@uni.lu

## ABSTRACT

Genome-wide association studies (GWAS) identify correlations between the genetic variants and an observable characteristic such as a disease. Previous works presented privacy-preserving distributed algorithms for a federation of genome data holders that spans multiple institutional and legislative domains to securely compute GWAS results. However, these algorithms have limited applicability, since they still require a centralized instance to decide whether GWAS results can be safely disclosed, which is in violation to privacy regulations, such as GDPR. In this work, we introduce GenDPR, a distributed middleware that leverages Trusted Execution Environments (TEEs) to securely determine a subset of the potential GWAS statistics that can be safely released. GenDPR achieves the same accuracy as centralized solutions, but requires transferring significantly less data because TEEs only exchange intermediary results but no genomes. Additionally, GenDPR can be configured to tolerate all-but-one honest-but-curious federation members colluding with the aim to expose genomes of correct members.

## 1 INTRODUCTION

Genome-Wide Association Studies (GWAS) are essential statistical analyses that aim at better understanding the correlations between genomic variations and observable traits or diseases, called phenotypes. GWAS play an important role in human health research and the development of precision/personalized medicine [10].

As the precision of GWAS increases with the size of the genomic dataset they study, biocenters are incentivized to perform studies collaboratively. A variety of cryptographic solutions have been proposed to compute GWAS statistics in a distributed and secure manner [7–9, 16, 21, 27, 46]. However, secure computation is not sufficient, because privacy can also be attacked from just observing the results of such a GWAS release. For instance, membership inference attacks [11, 23, 24, 40, 51] have been reported to identify whether an individual's known genome participated in a given study. Such a revelation would potentially lead to the identified individuals being discriminated [3], a risk that may well extend to their relatives [2].

To enforce privacy, once publicly available GWAS results are now only accessible under restrictive conditions [53] and regulations, such as the Health Insurance Portability and Accountability Act (HIPAA) [31] and the General Data Protection Regulation (GDPR) [33]. Both impede an exchange of personal data across borders [21, 34, 35]. Hence, to facilitate open research and the sharing of scientific findings, it is necessary to design privacy-preserving GWAS solutions that are compatible with existing regulations.

Two main techniques have been proposed to enforce privacy while openly releasing GWAS results: (i) differential privacy (DP) [20] and (ii) methods that bound the statistical inference power of an adversary [23, 40, 51, 55].

Differential privacy perturbs the results of a GWAS to obtain privacy at the cost of decreased accuracy. In contrast, methods that bound the adversary's statistical inference power aim to preserve high accuracy by performing several successive computations over the data to identify a subset of genomic variants over which statistics can be securely computed and safely released. More specifically, they ensure that the detection power of an adversary attempting membership inference attacks remains below a given threshold, which is derived from a tolerated false-positive rate [23, 40, 51]. Unfortunately, such computations require genomic data to be pooled in a central location [23, 36, 37, 40, 51].

This work strives for a federated GWAS protection mechanism that bounds the statistical inference power and that: (i) maintains the same privacy guarantees against external adversaries as one would obtain from centralized solutions; (ii) provides protection against possible collusions of up to all-but-one federation members; and (iii) keeps all genomic data inside the participating biocenters' premises.

---

This paper introduces Genome Distributed Private Release (GenDPR), a distributed middleware that achieves the above goals. Each federation member runs two subsystems of GenDPR. A non-trusted one (from the perspective of other federation members) that exclusively accesses the local genomic data of that member, and a subsystem that is trusted by all federation members to combine intermediary information and to identify the subset of variants that can be safely used in the subsequent secure GWAS computation. GenDPR leverages Trusted Execution Environments (TEEs) (e.g., Intel SGX enclaves), but works as well with other privacy-preserving schemes, such as fully homomorphic encryption [30].

Running GenDPR, federation members only exchange intermediate data, such as allele count vectors and local correlation metrics, instead of genomic variants files that can amount to several gigabytes and which likely also include irrelevant information for the task at hands. GenDPR significantly reduces the secure storage requirement of the TEE that will be elected to determine whether it is safe to release GWAS results. GenDPR outsources and communicates intermediate data exclusively in encrypted form and only to properly authenticated TEEs, as a release of such information would still enable some membership and inference attacks [36], albeit with a much reduced chance of success. We evaluate the performance of GenDPR considering several GWAS scenarios using up to 10,000 SNPs and 27,895 real genomes, which we compare against a centralized and non-collusion-tolerant approach.

The remainder of this paper is organized as follows. We discuss related works in Section 2 and provide necessary background in Section 3. Section 4 describes our system and threat models. We introduce GenDPR in Section 5 and further discusses how honest-but-curious collusions are tolerated. Next, Section 6 explains GenDPR's algorithm in greater detail. Section 7 presents GenDPR's results discussion, and show that GenDPR does not affect the accuracy of the tests used to identify the safe subset over which GWAS results can be released. Finally, Section 8 concludes this paper and directs to future work.

## 2 RELATED WORK
### 2.1 Privacy-Preserving Federated GWAS

Several solutions have been proposed to protect federation members' privacy-sensitive information during distributed computations of GWAS. These can be based on cryptographic mechanisms, such as Secure Multiparty Computation (SMC) [16, 25, 46] and Homomorphic Encryption (HE) [7, 26, 30], leveraged trusted execution environments (TEEs) [13, 14, 36, 37], or Differential Privacy (DP) [18, 28]. SMC allows parties to privately share their inputs for computing aggregate functions without the need for trusted third parties. HE operates entirely on encrypted data and reveals access to the final output only to authorized players holding the corresponding keys. DP perturbs local outputs by adding noise so that the probabilities that any individual's data was used or not differ by a limited amount.

Hybrid schemes, like SAFETY [39] or SCOTCH [15] combine HE for aggregation and TEEs for computing more complex statistics. Carpov and Tortech [12] compute the $\chi^2$ statistics inside an Intel SGX enclave using horizontal partitioning techniques to encode genomic data more efficiently. Kockan et al. [27] filters and compresses variant files (under the VCF file format) in a TEE for more efficient sharing. Bomai et al. [8] combine multi-key HE and TEE to enable secure sharing of genomic data from multiple institutions, which are later used by the SGX-enabled service to compute and answer GWAS queries.

TEEs leverage trusted processor areas. Their security can further be reinforced through memory oblivious and side-channel mitigation techniques (Section 2.3 discusses this topic further). GenDPR leverages Intel SGX [19] as TEE, but can easily be extended with side-channel defense mechanisms or ported to a different TEE [34].

### 2.2 Private GWAS Releases

In addition to protecting privacy-sensitive genomic information while GWAS results are produced, federation members must also ensure that revealed results carry no information that allow inferring the participation of individuals in studies, in particular in the case group, (e.g., by adversaries mounting membership inference attacks [23, 24, 40, 52]).

Statistical inference-based methods can be used to determine which data should be used to create safe releases. Homer et al.'s membership attack [23] leverages likelihood-ratio tests (LR-tests) to identify the presence of a particular individual in the case population. Wang et al. [51] proposed hypothesis test $T_r$ to ensure that the identification power of participants remains below a given threshold. Zhou et al. [55] proposed the $\Lambda$ metric for the same purpose. SecureGenome (SG) [41] combines several genome-oriented statistical verifications and a LR-test to select a subset of SNPs whose GWAS results are safe to be used. DyPS [36] builds upon SG and leverages a TEE to centrally select SNP subsets over which it is safe to release GWAS statistics that are computed in a federated and dynamic manner, i.e., as soon as new genomes become available.

DP can be used to protect the results of GWAS analyses by perturbing the final statistics [4, 43] or when sharing intermediate data during federated computation. However, care has to be taken to not disturb results to a degree where high-precision studies become infeasible [42, 43]. For this reason, centralized DP-based approaches often use central curators, which, given access to the genome information, compute how much noise still leads to safe and accurate results.

|  | SNP$_1$ | SNP$_2$ | ... | SNP$_L$ | Population |
|---|---|---|---|---|---|
| Individual $gen_1$ | 0 | 1 |  | 1 | Case |
| Individual $gen_2$ | 1 | 1 |  | 1 | Control |
| ⋮ |  |  |  | ⋮ | ⋮ |
| Individual $gen_N$ | 0 | 1 |  | 0 | Case |

Table 1: Collected and encoded data for GWAS.

In this paper, we avoid centralizing genome information, in particular because doing so raises legislative and data protection concerns. Instead, we propose GenDPR, as a distributed solution to assess the risk associated with using a particular subset of SNPs in a federated GWAS. In addition, GenDPR offers collusion-tolerance, which we argue is a further important aspect towards increased privacy guarantees.

## 2.3 TEE (Intel SGX) limitations

TEEs generally suffer from limited memory (128 MB) [12]. In response, SGX2 allows dynamic memory management and paging techniques to expand an enclave's memory to up to 4 GB [13]. Several TEEs experience side-channel attacks that can allow adversaries to exploit the memory access patterns of algorithms running inside the enclave to leak private information like genome data [13, 29, 36, 37]. Generic memory-oblivious solutions have been proposed to overcome this issue, such as path RAM (PRAM) [45], Oblivious RAM (ORAM) [22], and Oblivious B+ tree shuffling [49]. Specific memory-oblivious genomic data processing algorithms have also been developed [1, 13, 14, 29]. Data-oblivious approaches have a significant performance overhead [1, 29]. Adapting GenDPR to implement data-oblivious access patterns is a challenge that we leave as future work.

## 3 BACKGROUND

### 3.1 Genomics 101

Humans share almost 99% of the 3 billion nucleotide pairs contained in their genome. The remaining 1% are called genetic variations. The vast majority of such variations are Single Nucleotide Polymorphisms (SNPs), where one nucleotide is replaced by another one. Genomic variations indicate unique biological characteristics, such as disease dispositions. Genome-Wide Association Studies (GWAS) aim at revealing such correlations between variants and observable characteristics, which are also called phenotypes.

A genome-wide association study over $L$ SNPs encodes each genome using a binary value per SNP and indicates the population it belongs to. This encoding is illustrated in Table 1. The case population contains the individuals that have the phenotype of interest, while the control population contains the remaining individuals. For each individual, at a given genomic variation $SNP_l$ ($l \in \{0, \ldots L\}$), if its genome only contains the most common SNP (i.e., the major allele) it is then encoded using a 0, while the presence of the least common allele (i.e., the minor allele) is encoded with a 1.

GWAS routinely require the computation of intermediary tables, illustrated in Tables 2a and 2b, which are later used to compute GWAS statistics. In Table 2a, $N_i^{case}$ and $N_i^{control}$ stand for the count of the major/minor allele occurrence of the respective population. $N^{case}$, $N^{control}$, $N_0$, $N_1$, and $N_T$ are the sums of the columns and rows. In Table 2b, the $C_{--}^{l1,l2}$ are the pairwise allele counts, i.e., the number of occurrences of the minor/major allele combinations {00, 01, 10, 11} between two SNP positions $l_1$, $l_2$. An example of a test statistic is the Linkage Disequilibrium (LD), which identifies correlations between any two SNP positions $l_i, l_j$: $r^2 = \frac{(C_{00}^{i,j} \cdot C_{11}^{i,j} - C_{01}^{i,j} \cdot C_{10}^{i,j})^2}{C_{0-}^{i,j} \cdot C_{1-}^{i,j} \cdot C_{-0}^{i,j} \cdot C_{-1}^{i,j}}$. Additionally, $\chi^2$ tests are computed to measure the association of a SNP with the phenotype of interest, defined as $\chi^2 = \frac{(N_i^{case} - N_i^{control})^2}{N_i^{control}}$. $P$-values on $\chi^2 < 10^{-8}$ identifies a strong association with a particular phenotype [5]. The SNPs with the smallest $p$-values are the most significant (ranked) SNPs of a GWAS.

To increase the confidence in GWAS findings [32] and avoid bias (e.g., through under-representing some populations, like Asians, Africans and Latin Americans compared to European and North American individuals) [44], biocenters move towards global-scale GWAS [6, 48]. Under these large scale settings, federations of biocenters assemble a global dataset comprised of multiple smaller datasets sampled from possibly geographically distant populations [27, 39].

| SNP$_l$ |  | Case | Control | Total |
|---|---|---|---|---|
|  | 0 (major) | $N_0^{case}$ | $N_0^{control}$ | $N_0$ |
|  | 1 (minor) | $N_1^{case}$ | $N_1^{control}$ | $N_1$ |
|  | Total | $N^{case}$ | $N^{control}$ | $N_T$ |

(a) A singlewise contingency table.

|  |  | SNP$_{l_2}$ | | Total |
|---|---|---|---|---|
|  |  | 0 | 1 |  |
| SNP$_{l_1}$ | 0 | $C_{00}^{l_1,l_2}$ | $C_{01}^{l_1,l_2}$ | $C_{0-}^{l_1,l_2}$ |
|  | 1 | $C_{10}^{l_1,l_2}$ | $C_{11}^{l_1 l_2}$ | $C_{1-}^{l_1,l_2}$ |
|  | Total | $C_{-0}^{l_1,l_2}$ | $C_{-1}^{l_1,l_2}$ | $2N^{pop}$ |

(b) A pairwise contingency table.

Table 2: GWAS contingency tables.

## 3.2 Statistical Tests for Safe Release

Like SG [40] and DyPS [36], GenDPR leverages the likelihood-ratio test to assess the privacy risks associated with considering particular SNPs in a GWAS statistics. We briefly recall GWAS statistics and the principle of the likelihood-ratio test in the following.

### 3.2.1 Minor Allele Frequencies (MAF).
SNP positions with rare minor alleles (e.g., MAF < 0.05) form characteristic outliers, which can be used by adversaries to deduce membership [51, 55]. SNPs with very rare alleles are therefore not considered in a privacy-preserving GWAS whose result should be publicly available.

### 3.2.2 Linkage Disequilibrium (LD).
High linkage disequilibrium (e.g., $p$-value on $r^2 < 10^{-5}$) indicates highly connected SNPs. Such information can be leveraged to attack individuals by using the association levels among SNPs, as shown in [40, 55]. SNP positions in high LD should therefore be removed as well from public releases.

### 3.2.3 Likelihood-Ratio Test (LR-test).
MAF and LD ensure the independence of retained SNPs, which is a requirement for correctly conducting SG's LR-test. SG's null hypothesis corresponds to the situation where an individual does not belong to the case population, while under the alternate hypothesis the individual is part of that population. The LR metric is expressed as follows (adapted from [40]):

$$LR = \sum_{l=1}^{L} \left[ x_{n,l} \log \frac{\hat{p}_l}{p_l} + (1 - x_{n,l}) \log \frac{1 - \hat{p}_l}{1 - p_l} \right], \quad (1)$$

where $L$ is the number of SNPs tested, $x_{n,l}$ is the allele information at SNP position $l$ of individual $n$, $p_l$ is the frequency of SNP position $l$ in the reference set and $\hat{p}_l$ is the frequency of SNP $l$ in the case population. SG's authors empirically demonstrate that the LR-test is more powerful than previous metrics, including the one used in Homer et al.'s attack [23].

The LR-test can be used to find a subset of SNPs in $L$ that can have their statistics released while maintaining the power for detecting individuals in a study below a configured detection and false-positive rates [41]. Thus, GWAS statistics released over the retained SNPs enforce privacy [36, 37, 40]. Threshold parameters, i.e., MAF and LD cut-offs, and the confidence levels used for the LR-test are configured according to the privacy guarantees one desires to achieve [41].

## 4 MODELS AND OBJECTIVES

We illustrate GenDPR's system and threat models in Fig. 1 and describe them in the following.

*System model.* We consider a federation composed of $G$ members, which we also call genome data owners (GDO):

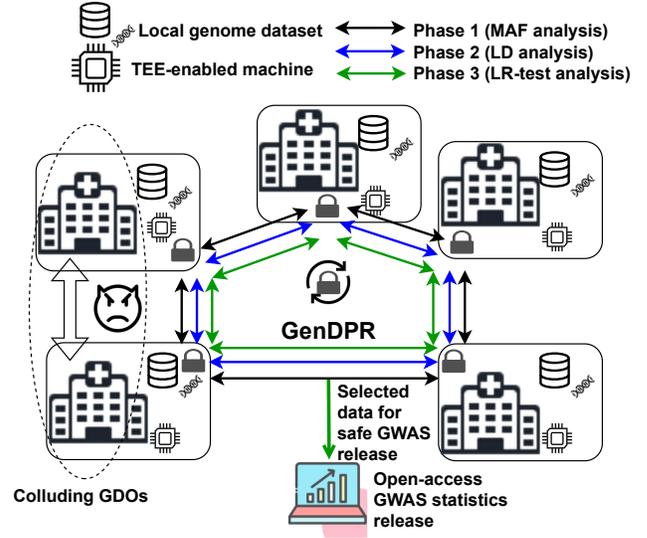

Figure 1: GenDPR's system and threat model.

$GDO_1, \ldots, GDO_G$. Each GDO is entrusted with genomic information from patients and authorized to use this information in genome-wide association studies. On their premises, federation members maintain a database with genomes and a TEE-enabled server to perform operations over this data, whose TEE is mutually trusted by that member and all others. Remote attestation ensures authenticity of the trusted part of GenDPR that runs in the TEE. Additionally, a TEE data-sealing mechanism is used to store data persistently outside the TEE. Sealed data can only be encrypted/decrypted by the enclave using its private key. We further assume that appropriate countermeasures are in place to mitigate potential limitations of a particular TEE implementation. Members have access to a common public genome dataset (e.g., from the 1000 genomes project [17] or dbGaP [50]), which they use as reference for the LR-test.

*Threat model.* Like previous works on secure GWAS release [36–39, 54], we assume adversaries are capable of mounting membership attacks by observing released GWAS statistics and metadata (see Section 3). In a membership attack, the adversary owns a victim's genotype sequence ($gen_{victim}$) and has access to a reference population with an allele distribution similar to the one of the case population used in the study [23, 41, 51]. The adversary's goal is to infer participation of $gen_{victim}$ in the GWAS, which can result in severe privacy implications. For example, inferring that a particular individual belongs to the case population can be unethically used (e.g., by insurance companies refusing coverage).

In addition, we assume that up to $f$ federation members might be faulty (e.g., as a result of compromise) and collude with other compromised members to mount membership attacks. Colluding allows members to exchange their knowledge, which increases their chances of mounting membership attacks successfully (as reported in Pascoal et al. [36]). We allow $f$ to become as large as $G - 1$, but will make no further liveness guarantees once federation members become non-responsive. We leave the addition of measures to guarantee liveness despite faults as future work. Likewise, we do consider leakage of genome information from the member's premises an orthogonal problem.

While we assume that federation members might be honest-but-curious and collude in their attacks, we assume that the integrity and confidentiality of each member's TEE remains intact. Moreover, we assume that the trusted part of GenDPR is able to detect whether a federation member has tampered with the genome data and its accuracy (e.g., by checking the authenticity of signed VCF files and all exchanged data).

*Objectives.* Given a desired starting set of SNP positions $L_{des}$, GenDPR returns a reduced set $L_{safe} \subseteq L_{des}$ of SNPs that are safe to be considered in a GWAS, which the federation computes after this check using one of the existing secure and privacy-preserving federated GWAS approaches [8, 27, 36–39].

We aim at securing the privacy of individuals that have entrusted a correct federation member (GDO) with their genomes even when the federation releases GWAS results or when other members are honest-but-curious or get compromised.

Under the above assumptions, we show that as long as no TEE crashes, GenDPR correctly produces a selection of SNP positions ($L_{safe}$) that are safe to be considered for actual GWAS computation while protecting the privacy of individuals, even if up to $f \leq G - 1$ federation members collude. Thus, GenDPR ensures that the GWAS federation properly considers the risks of including genetic variations that might compromise the privacy of its members.

To remain compliant with regulations, such as the GDPR, no raw genomic information gets exchanged and all communication of intermediate results remains encrypted and exclusively among the TEEs. Despite that, GenDPR mimics the same output decision as if a centralized privacy-protection mechanism is in place.

## 5 GENDPR
### 5.1 Overview
GenDPR coordinates multiple TEEs at the GDO's premises, allowing them to jointly conduct a particular GWAS aiming at releasing statistics over $L_{des}$ SNP positions with specific MAF, LD and LR-test cutoff parameters. GenDPR proceeds with a randomly elected leader GDO. The TEEs of non-leader GDOs produce the required intermediate values and compute the GWAS on their selected local subsets. The leader's

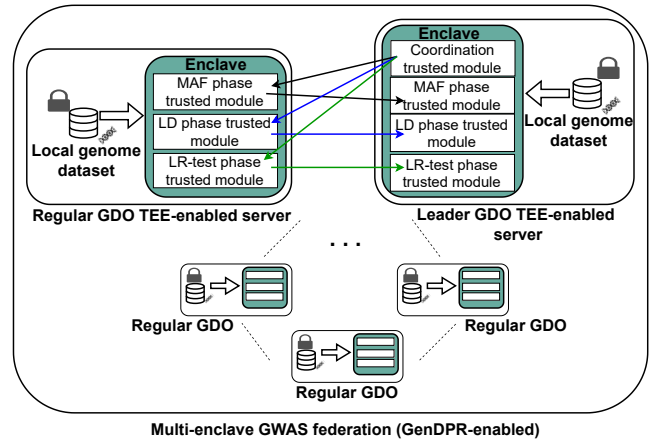

Figure 2: GenDPR's architecture components.

TEE coordinates this computation and aggregates the intermediate results it receives from other federation members, but also produces such results for the local dataset.

Encrypted local genome datasets are used to feed local enclaves so that each GDO enclave can produce and outsource genomic intermediate data as requested by the leader according to the current phase of the protocol. Since TEEs are mutually authenticated to each other, they trust the code and results produced in this outsourcing step. Any communication between federation members is encrypted and happens only between TEEs. That is, the sending TEE will encrypt the information such that only the receiving TEEs can encrypt it. In particular, GDOs agree on keys and other credentials during the remote attestation phase to connect the trustchain from boot to communication. Fig. 2 illustrates GenDPR's architecture.

### 5.2 Workflow
We present an overview of GenDPR's workflow in Fig. 3. GenDPR follows three consecutive phases where GDOs compute and outsource different intermediate computation results. One of the GDOs is randomly chosen as coordinator of the protocol and aggregator of inputs produced by the other GDOs. In particular, at the beginning of a GWAS and before the start of its distributed computation, GenDPR initiates two essential *pre-processing* tasks: First, the leader TEE selection, which consists of randomly choosing one of the registered enclaves of participants of the federation. Second, the computation of summary statistics (e.g., allele count vectors of each GDO over the original SNP-set $L_{des}$ over which the GWAS would ideally be computed). As part of this second step, GDOs locally compute $N^{case}_{l}{}_l$ for each $l \in L_{des}$. Such a vector is identified as *caseLocalCounts*$[L_{des}]_g$ of size $L_{des}$ and is sent by each GDO $g$ to the leader GDO's enclave for aggregation. The GDOs also share the number of individuals in their local case population ($N^{case}_g$). All of the above information is encrypted and linked to the boot of the involved TEEs, so that only a properly authenticated enclave can decrypt them. With this information in place, GenDPR finds the set of secure-to-be-used SNPs in the following three phases.

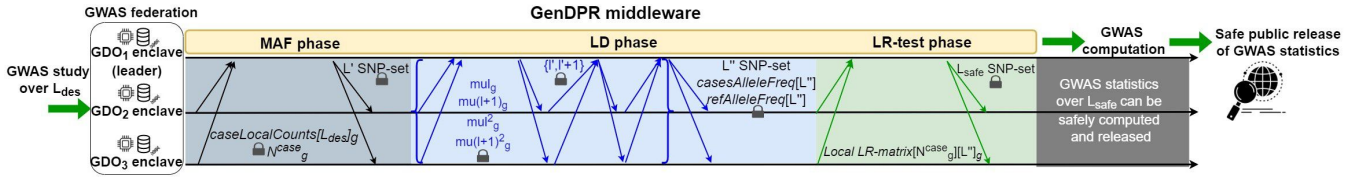

Figure 3: GenDPR's workflow.

### 5.3 Phase 1/3: MAF Analysis

First, the leader enclave locally computes the allele counts vector of the reference population *referenceLocalCounts*[$L_{des}$], which contains $L_{des}$ entries that are each computed over $N^{reference}$ genomes. Then, after receiving the encrypted *caseLocalCounts*[$L_{des}$]$_g$ and $N^{case}_g$ from each GDO, the leader enclave decrypts and starts MAF verification. In particular, the leader enclave sums all $N^{case}_g$ received from the GDOs with $N^{reference}$ into $N_T$. Then, the leader GDO goes over the received inputs to calculate the allele counts of SNPs in both populations (case and reference) and then computes the global MAF of each SNP. More specifically, for each $l$ in the original SNP-set $L_{des}$ and for each GDO's $g$ allele counts vector, the leader GDO computes *totalGlobalCounts*[$l$] as the sum of *caseLocalCounts*[$l$]$_g$ and *referenceLocalCounts*[$l$]. The aggregated result is then divided by $N_T$ to obtain the MAF for SNP $l$, i.e., *globalAlleleFreq*[$l$] is computed as *totalGlobalCounts*[$l$]/$N_T$. Finally, the leader checks if $MAF_l < MAF_{cutoff}$. If so, SNP $l$ is removed and will not further be considered for the current release. Phase 1 removes rare MAF SNP positions without requiring to outsource the actual genomes from the GDOs. At the end of this phase, the leader GDO broadcasts the list of retained SNPs $L' \in L_{des}$ to all federation members, which is then further reduced in Phase 2.

### 5.4 Phase 2/3: LD Analysis

The second phase consists in executing the Linkage Disequilibrium (LD) verification over the retained SNPs $L'$. It ensures all released SNPs will be independent from each other. To compute LD, allele information between two SNPs needs to be pooled. This is easily achievable in a centralized TEE-based architecture with local access to all genomes. However, in GenDPR, we aim to keep genomes distributed at their respective GDO, which prevents us from pooling allele sequences for LD computation.

One could naïvely let GDOs conduct the LD analysis over their local data and share the locally retained SNPs. However, this approach would lead to inaccurate selection since each GDO's local data does not incorporate the heterogeneous distribution of genomes among the GDOs.

GenDPR therefore employs the following adaptations for removing SNPs in LD. When computing the LD between every pair of SNP $l$ and $l + 1 \in L'$, local allele sequences of individuals are pooled to compute correlation statistics. For that, each GDO $g$'s enclave locally produces and outsources the following correlation statistics over their genomes: $\mu_{l_g}\ +=\ \text{SNP}_{l_g}$, $\mu_{l+1_g}\ +=\ \text{SNP}_{l+1_g}$, $\mu_{(l,l+1)_g}\ +=\ \text{SNP}_{l_g} * \text{SNP}_{l+1_g}$, $\mu_{l_g^2}\ +=\ \text{SNP}_{l_g} * \text{SNP}_{l_g}$, $\mu_{(l+1)_g^2}\ +=\ \text{SNP}_{l+1_g} * \text{SNP}_{l+1_g}$, and $N_T$ (acquired during the previous phase). The leader GDO computes the same correlation statistics over the reference set.

Upon receiving the member's correlation statistics, the leader enclave aggregates GDO inputs with the correlation metrics obtained over its local and the reference set. This way, GenDPR absorbs the correlation statistics from each GDO so that the aggregated correlation metrics reflect the global genome distribution of the federation for a proper computation of LD. After that, the leader enclave can proceed with the computation of the $p$-value on the $r^2$ test to measure how much the two SNPs are correlated. If $\text{LD}_{(l,l+1)} < LD_{cutoff}$, then SNPs $l$ and $l + 1$ are dependent, and cannot both be retained. GenDPR keeps the higher ranked (in terms of $p$-value on $\chi^2$, recall Section 3). The leader iterates over this step at most $(L')^2$ times, which would happen if all pairs of SNPs were independent, which remains an unlikely common event in case of the human genome [5, 10]. The result of this phase is a further reduced list of SNPs $L''$, which the leader broadcasts to all members to start phase 3.

### 5.5 Phase 3/3: LR-test Analysis

Similar to the previous phase, the allele information of each participant is needed (i.e., $x_{n,l}$ in Equation 1) to perform LR-test verification. Existing solutions rely on a centralized enclave that collects all required data. On the other hand, GenDPR overcomes such constraint by demanding each GDO to compute and outsource their local LR-matrices. The local LR-matrix consists of the LR values (recall Equation 1) for each SNP $l$ and of the allele value of individual $n$ at SNP position $l$ represented by $x_{n,l}$ in each GDO dataset. However, GDOs cannot correctly compute these matrices just from their local genome dataset, since this would lead to incorrect conclusions. Indeed, using local frequencies only would lead to wrong LR-matrices because the LR-test needs to be drawn considering aggregate SNP frequencies of the whole population (i.e., genomes belonging to all GDOs). Therefore, allele frequencies over the full cohort are needed so that each GDO can accurately compute their local LR-matrix.

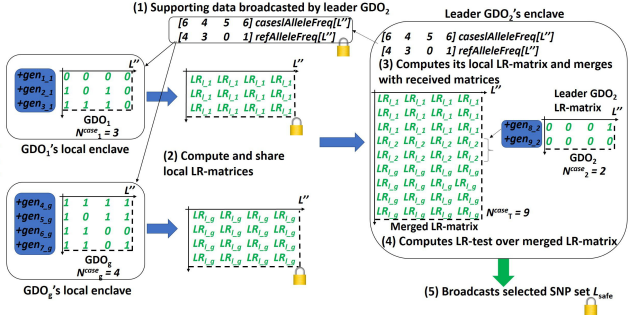

Figure 4: GenDPR distributed LR-test phase scheme.

The complete scheme enforced by GenDPR for the distributed LR-test evaluation is illustrated in Fig. 4, where we consider that $GDO_2$ has been selected as leader. In Step (1), the leader broadcasts the allele frequencies vector of the case and reference populations over the retained SNPs $L''$ (note that these vectors are already available inside the leader enclave since the MAF phase). Therefore, the $casesAlleleFreq[L'']$ and $refAlleleFreq[L'']$ vectors, both of size $L''$, are shared with all GDOs. These vectors represent $p_l$ and $\hat{p}_l$ of Equation 1, respectively. In Step (2), after the reception of the allele frequencies vectors, each GDO ($GDO_1$ to $GDO_g$) can correctly build their LR-matrices since the received vectors encompass the frequencies over the complete cohort of participating genomes. Thus, their local LR-matrices can be correctly computed. After completion, GDOs encrypt and send their local LR-matrices to the leader GDO. In Step (3), upon reception of the GDOs' LR-matrices, the leader first computes its local LR-matrix, and then merges all matrices received. As a result, creating a larger LR-matrix that encompasses all GDOs data. This matrix is used throughout the LR-test verification performed in Step (4) inside the leader's enclave. This verification consists of empirically checking several subsets of SNPs in $L''$ that satisfies the conditions presented in Section 3.2. When the LR-test ends, the leader enclave has identified a new subset of SNPs $L_{safe} \in L''$, which is encrypted and broadcast to the members of the federation (Step (5)). The list of SNPs in $L_{safe}$ can be safely used for the computation and release of the GWAS. We present the pseudo-code of GenDPR's protocol in Section 6.

Additionally, we note that GenDPR could be combined with Differential Privacy (DP) [20] to increase the data utility of releases. The SNPs in $L_{safe}$ can be released in a noise-free manner (i.e., without any data perturbation), while statistics over SNP positions in the complement $L_{des} \setminus L_{safe}$ could be released, but with DP-perturbation. Such a hybrid scheme would allow the release of GWAS statistics over all desired SNP positions ($L_{des}$) in a privacy-preserving manner.

### 5.6 Tolerating Honest-but-Curious Collusions

To protect the GWAS federation against collusion among GDOs, GenDPR's leader enclave needs to certify that the outcome of the private analysis is valid for the cases where up to $f \leq G-1$ colluding GDOs attempt to attack the honest ones. For this purpose, GenDPR employs a collusion-tolerant algorithm. For each phase of GenDPR's pipeline discussed above, GenDPR generates $\binom{G}{G-f}$ combinations of intermediate results received from the GDOs to simulate the case where $f$ GDOs would launch an attack. Each of these combinations has a unique identifier and goes through the various phases of GenDPR, which identify a list of safe SNPs. At the end of each phase, GenDPR computes the intersection of the SNPs chosen for each combination, thus preventing any $f$ GDOs to compromise the data of honest GDOs. Let us discuss an example for Phase 3 (the most complex one).

During the LR-test phase, the leader enclave generates and provides a unique id, and broadcasts $\binom{G}{G-f}$ allele frequency vectors over $L''$ SNPs selected in the previous LD analysis phase. The leader then receives $\binom{G}{G-f}$ local matrices (each one computed using its corresponding frequency vector) from each GDO. Each combination of sub-matrices forms a unique merged matrix that is used for the actual LR-test evaluation inside the leader enclave. As a result, GenDPR collects several lists of selected SNPs ($L_{safe}$), i.e., one for each LR-test completed over each combination of matrices. Finally, the leader enclave computes the intersection among the lists of SNPs, to finally output only the intersected SNPs, that were mutually labeled as safe in every combination. This way, GenDPR certifies that no combination of genome data can be isolated and become vulnerable to colluding GDOs. GenDPR can also adhere to a more conservative approach assuming all possibilities of collusions instead of considering a static $f$, i.e., $f = \{1, ..., G-1\}$. GenDPR would then perform evaluations over $\sum_{f=1}^{f=G-1} \binom{G}{G-f}$. As one would expect, this scheme demands GenDPR to execute extra rounds of computations, which in practice can be efficiently conducted in parallel inside the leader enclave as it already stores all necessary data.

# 6 GENDPR PSEUDO-CODE

Algorithm 1 shows the pseudocode of GenDPR. It reflects the behavior and steps explained above. We refrained from describing the standard methods we use for encypting operations during the workflow and focus here on the rationale of the algorithm.

In Line 6, GenDPR randomly selects a leader GDO among the federation members. Then, in line 9, the leader GDO starts computing its local GWAS summary statistics. The same computation is also applied to the genomes of the reference set. From that moment on, the leader receives summary statistics of the other GDOs that are locally computed when the federation agrees on starting a study. After collecting the other GDOs' intermediate data, the leader GDO proceeds with MAF analysis by first aggregating local counts over the original SNP-set $L_{des}$. It does the same for calculating the total number of individuals in the federation. Then, the leader GDO finally computes the MAF of each SNP and checks the MAF cut-off, keeping only SNP positions with MAF above or equal to the MAF cut-off ($MAF_{cutoff}$). These steps are described in lines 10–24. At the end of this analysis, the leader GDO has acquired a new SNP-subset $L'$ consisting of the list of SNPs that survived this phase. Such a list of SNPs is broadcast to all GDOs (line 25).

Next, the leader GDO initiates the LD analysis after receiving the correlation metrics of each GDO of pairwise SNP combinations in $L'$. In particular, the LD verification algorithm (from lines 26 to 55) aggregates local correlation statistics from each GDO and the ones corresponding to the reference set for SNPs pair $l$ and $l + 1$. In addition, the leader GDO computes allele frequencies over $L''$ for the reference and global population (note that is achievable using the allele counts shared in the MAF analysis) that is further aggregated with the correlation metrics of all GDOS. After aggregation, the leader GDO calculates the $p$-value for the correlation between the two SNPs. If SNPs are high-correlated, i.e., $p$-value below the LD cut-off ($LD_{cutoff}$), the leader GDO keeps the most ranked SNP and proceeds the loop. SNPs that do not present a high pairwise correlation with others are retained in $L''$, which is also broadcast at the end of this phase. Finally, the allele frequencies vectors ($casesAlleleFreq[L'']$ and $refAlleleFreq[L'']$) are broadcast to the GDOs (line 56).

Last, the leader GDO performs the LR-test to find the final list of safe SNPs. This verification starts in line 58, where the leader GDO receives the local LR-matrices from each GDO that are locally computed by each GDO using $casesAlleleFreq[L'']$ and $refAlleleFreq[L'']$ shared in the

**Algorithm 1** GenDPR's full workflow pseudo-code

1: **procedure** GenDPR(GWAS $s$, $G$ set of GDOs, original SNP set $L_{des}$ of $s$, $\alpha$, $\beta$, $ref\_population$)
2:    **Inputs:** (1) Local allele counts vector from GDOs of size $L_{des}$; (2) Local statistics of $SNP_l$ and $SNP_{l+1}$; (3) Local LR-matrix each of size $N^{case}{}_g$ x $L''$
3:    **Outputs:** (1) Selected SNP subset $L'$; (2) Selected SNP subset $L''$; (3) Selected SNP subset $L_{safe}$, which can be used to create private GWAS release
4:    **Uses:** $randomLeaderSelection(G)$: select and returns a random GDO $g \in G$ to be considered as leader; $startLocalComputations()$: computes local statistics of GDO; $computeR^2(\mu_l, \mu_{l+1}, \mu_{(l,l+1)}, \mu_{l^2}, \mu_{(l+1)^2}, N_T)$: returns $p$-value on $r^2$ between SNPs $l$ and $l + 1$; $getMostRanked(l, l + 1, s)$: returns index of most ranked SNP ($p$-value on $\chi^2$ of study $s$; $LRtest(LRMatrix, \alpha, \beta)$: returns a set of SNPs that keeps individuals identification power below given threshold
5:
6:    $leader_{gdo}$ = $randomLeaderSelection(G)$ //randomly selects a leader in $G$
7:    $leader_{gdo}.startLocalComputations()$ // computes leader and GDOs local allele statistics
8:    $leader_{gdo}.listenToInputs()$ // collects intermediate data from other GDOs
9:
10:    (Phase 1) //MAF analysis
11:    **for** $g$ in $G$ **do** //retrieves local allele counts vector from each GDO
12:       $N_T$ += $N^{case}{}_g$
13:    **end for**
14:    $N_T$ += $N^{reference}$
15:    **for** SNP $l$ in $L_{des}$ **do**
16:       **for** $g$ in $G$ **do** //retrieves local allele counts vector from each GDO
17:          $totalGlobalCounts[l] = caseLocalCounts[l]_g + referenceLocalCounts[l]$
18:       **end for**
19:       $globalAlleleFreq[l] = totalGlobalCounts[l]/N_T$
20:       **if** $globalAlleleFreq[l] < MAF_{cutoff}$ **then** //SNP $l$ cannot be retained
21:          **continue**
22:       **else**
23:          $L'.push(l)$
24:       **end if**
25:    **end for**
26:    $leader_{gdo}.broadcast(L')$ // leader GDO broadcast message
27:
28:    (Phase 2) //LD analysis
29:    $last_{index} = L'[-1]$ // get index of the last SNP in $L'$
30:    $aux_{index} = L'[0]$ // get index of the first SNP in $L'$
31:    **while** $aux_{index}$ != $last_{index}$ **do** // starts greedy algorithm for LD computation
32:       **for** SNP $l$ in $L'$ **do**
33:          $leader_{gdo}.listenToInputs()$ // collects intermediate correlation statistics from GDOs
34:          $leader_{gdo}.startLocalComputations()$ // computes leader local correlation statistics
35:          **for** $g$ in $G$ **do** //retrieves local LD statistics for $SNP_l$ and $SNP_{l+1}$ from each GDO
36:             $\mu_l$ += $\mu_{l_g}$
37:             $\mu_{l+1}$ += $\mu_{l+1_g}$
38:             $\mu_{(l,l+1)}$ += $\mu_{(l,l+1)_g}$
39:             $\mu_{l^2}$ += $\mu_{l_g^2}$
40:             $\mu_{(l+1)^2}$ += $\mu_{(l+1)_g^2}$
41:          **end for**
42:          $\mu_l$ += $\mu_{l_{ref}}$
43:          $\mu_{l+1}$ += $\mu_{l+1_{ref}}$
44:          $\mu_{(l,l+1)}$ += $\mu_{(l,l+1)_{ref}}$
45:          $\mu_{l^2}$ += $\mu_{l^2_{ref}}$
46:          $\mu_{(l+1)^2}$ += $\mu_{(l+1)^2_{ref}}$
47:          $pval = computeR^2(\mu_l, \mu_{l+1}, \mu_{(l,l+1)}, \mu_{l^2}, \mu_{(l+1)^2}, N_T)$
48:          **if** $pval > LD_{cutoff}$ **then** //independent SNPs
49:             $aux_{index} = l + 1$
50:             **continue**
51:          **else** //dependent SNPs, keep most ranked one
52:             $l_{index} = getMostRanked(l, l + 1, s)$
53:             $L''.push(l_{index})$
54:          **end if**
55:       **end for**
56:       $aux_{index} = l + 1$
57:    **end while**
58:    $leader_{gdo}.broadcast(L'', casesAlleleFreq[L''], refAlleleFreq[L''])$ // leader GDO broadcast message
59:
60:    (Phase 3) //LR-test analysis
61:    $leader_{gdo}.listenToInputs()$ // collects local LR-matrices from GDOs
62:    $leader_{gdo}.startLocalComputations()$ // computes leader local LR-matrix
63:    **for** $g$ in $G$ **do** //retrieves and concatenates local LR-matrix from each GDO
64:       **for** SNP $l$ in $L''$ **do**
65:          $FullLRMatrix[l]$ += $LRmatrix_g[l]$
66:       **end for**
67:    **end for**
68:    $L_{safe} = LRtest(FullLRMatrix, \alpha, \beta)$ //runs LR-test analysis over full matrix
69:    **return** $L_{safe}$ //final subset of SNPs for safe GWAS $s$ release
70: **end procedure**

previous phase. Upon the reception of the local LR-matrices, the leader GDO loops over $L''$ to merge all received LR-matrices with its local matrix (lines 60 to 64) Next, in line 65, the leader GDO runs the LR-test function over the merged matrix that empirically finds a subset $L_{safe} \in L''$ of which releases over these SNPs do allow membership inference attacks to succeed. Finally, the leader GDO broadcasts $L_{safe}$ SNP-set list in line 66.

### 6.1 Collusion-tolerant GenDPR pseudo-code

In the following, we present in more detail the extensions to enable collusion-tolerant GenDPR. To avoid repeating the full GenDPR algorithm, we reuse Algorithm 1 and explain the required modifications to accommodate collusion-tolerance. To tolerate collusions, GenDPR needs to execute the analysis over each combination of data that can be actually isolated by colluding GDOs to mount membership attacks against honest GDOs.

To that extent, after retrieving intermediate data from each GDO in each phase, GenDPR forms $\binom{G}{G-f}$ combination with the received inputs to simulate the fraction of data that could be isolated by the colluding GDOs depending on $f$. Therefore the original set $G$, consisting of $g$ GDOs, becomes a new set of combinations of GDOs so that the verification can be computed for every possible combination of GDOs data. We call such a set $combGDOSet = combineGDOS(G)$, where $combineGDO(G)$ is a function that receives the set of GDOs $G$ and outputs a new set consisting of $\binom{G}{G-f}$ combinations. As a result, the GDOs data are collected according to each combination in $combGDOSet$. For instance, the loop for MAF analysis in Line 10 of Algorithm 1 is performed over $combGDOSet = \binom{G}{G-f}$ instead of $G$. The same behavior is applied to the other phases of GenDPR's protocol. Namely, in Line 33 for the LD analysis and Line 60 for the LR-test.

GenDPR further needs to keep a data structure to store the list of selected SNPs of each iteration. This is needed so that GenDPR can compute the intersection of SNPs selected as safe in all combinations. In fact, at the end of each phase, only SNPs present in all lists are broadcast to the federation because they are safe independently of the presence of colluders.

For example, considering the MAF phase, GenDPR appends each $L'$ to a new data structure called $L'\_ListSet$ after Line 23. Once the loop over $combGDOSet$ ends, GenDPR computes $finalL' = getIntersection(L'\_ListSet)$, where $getIntersection(.)$ is a function that receives a set of SNP lists and returns a list of SNPs mutually chosen in all combinations. The SNPs in $finalL'$ guarantees that no combination of intermediate results leveraged by colluding parties can be used to launch successful membership attacks.

The method to compute the intersection of SNPs is performed at the end of each phase, before data is broadcast by the leader GDO. More specifically, $getIntersection(L)$ finds the SNPs intersection over the list output for each iteration. It is executed before Line 25 for MAF analysis, Line 56 for the LD phase and before Line 66 after the LR-test verification, and then when acquiring the final intersected list of SNPs $L_{safe}$, which can be safely used in the GWAS computation whose results should be released.

## 7 EXPERIMENTAL EVALUATION

We implemented GenDPR in C/C++ using the Graphene SGX library [47] and evaluated its performance on an Intel i7-8650U processor with 16 GB RAM, running Ubuntu 18.04. For the experiments we used 27,895 real genomes from the phy001039.v1.p1 dbGaP dataset that was collected for an Age-Related Macular Degeneration study [50]. The dataset contains 14,860 case and 13,035 control genomes. We used the control population as reference for the LR-test. We have divided genomes equally among federation members and adopt suggested SecureGenome's [41] settings for assessing privacy — 0.05 MAF cut-off, $10^{-5}$ LD cut-off, 0.1 false-positive rate and 0.9 identification power threshold. We encrypt all exchanged data using AES 256. In our experiments, we vary the number of federation members (GDOs) between 2 and 7, and use between 1,000 and 10,000 SNP positions. We report the averages over 5 repetitions. We also compared GenDPR to a centralized approach that runs SecureGenome inside a centralized TEE enclave, which we use as baseline.

Table 3: GenDPR's average resource utilization.

| Configuration | CPU | Memory |
| --- | --- | --- |
| 2 GDOs / 1,000 SNPs | < 1% | 2,068 KB |
| 2 GDOs / 10,000 SNPs | < 1% | 2,164 KB |
| 3 GDOs / 1,000 SNPs | < 1% | 2,068 KB |
| 3 GDOs / 10,000 SNPs | < 1% | 2,172 KB |
| 5 GDOs / 1,000 SNPs | < 1% | 2,074 KB |
| 5 GDOs / 10,000 SNPs | < 1% | 2,148 KB |
| 7 GDOs / 1,000 SNPs | < 1% | 2,052 KB |
| 7 GDOs / 10,000 SNPs | < 1% | 2,180 KB |

### 7.1 Bandwidth, Memory and CPU Usage

Table 4 shows GenDPR's average resource demand in different configurations. As can be seen, all scenarios use less than 1% of the CPU and consume less than 2 MB on average of memory inside federation members' TEE. GDOs exchange vectors of integers that require 32 bits for each SNP in the original dataset $L_{des}$. Hence, the overall size of data that needs to be exchanged is $4 \cdot L_{des}$ Bytes, which increases by approximately 30% after encryption due to padding. With GenDPR, GDOs do not need to outsource genome sequences, which saves $2 \cdot L_{des}$ bits for every genome, i.e., $2 \cdot L_{des} \cdot N_T$ bits in total. Notice that the data that need to be exchanged in subsequent steps decreases as they successively operate on a decreasing number of retained SNPs. Indeed, for the LR-test phase, each GDO shares smaller data, i.e., over $L'' \cdot N^{case}_g$, which is an order of magnitude smaller than complete genome sequences. In summary, we see that GenDPR's performance scales well with an increasing number of GDOs and SNPs considered and that it remains well within the resource limitations of today's TEEs. Notice also that with additional GDOs, the resource demand remains low because the computation of the LR-test is distributed among them (i.e., they operate on smaller local datasets).

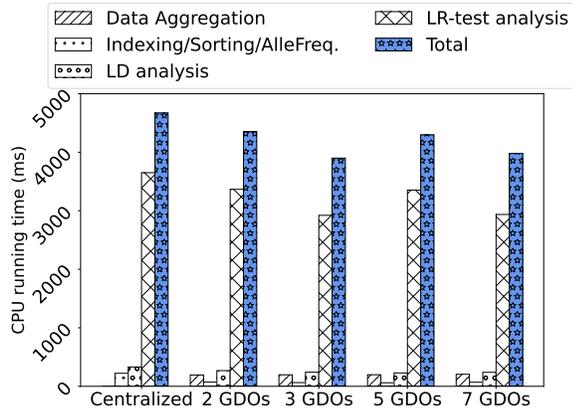

(a) 7,430 genomes / 1,000 SNPs.

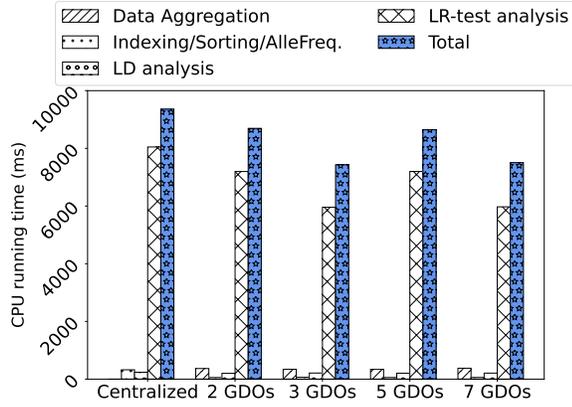

(b) 14,860 genomes / 1,000 SNPs.

Figure 5: Running time comparison (1,000 SNPs).

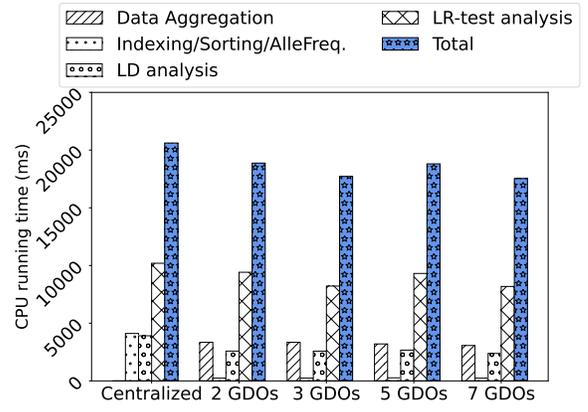

(a) 7,430 genomes / 10,000 SNPs.

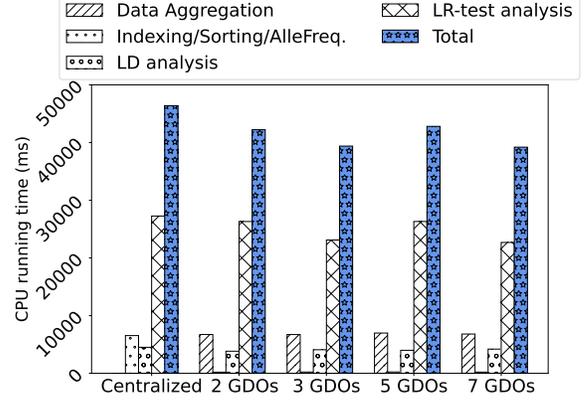

(b) 14,860 genomes / 10,000 SNPs.

Figure 6: Running time comparison (10,000 SNPs).

### 7.2 Running Time

Figures 5 and 6 report GenDPR's running time compared to the centralized baseline for each task performed during each phase over several possible GWAS settings. First, we can notice that even though it does not require any data aggregation tasks, the centralized solution is not noticeable faster than GenDPR. In particular, the running times of both approaches directly depend on the size of the data that need to be evaluated. Comparing Figure 5a with 5b, and Figure 6a with 6b, we can notice that increasing the number of considered genomes or SNPs significantly increases the running time of both approaches. Therefore, we claim that GenDPR is scalable since doubling the number of genomes considered at first (7,430) or considering 10 times more SNPs in a study have not rendered GenDPR unusable. Overall, GenDPR terminates under reasonable delays.

Moreover, one can see that increasing the number of GDOs actually decreases the running time of the protocol since the computational tasks are distributed among members, which demonstrates a performance improvement over the centralized baseline. In contrast, the centralized version cannot take advantage of such a feature, and therefore needs to process all the data at once. Hence, we claim that GenDPR also benefits from the workload distribution achieved thanks to its distributed protocol. An interesting phenomenon is that although the scenario with 5 GDOs presented a longer running time when compared to the scenarios with 3 and 7 GDOs, it takes approximately as long as the scenario with 2 GDOs. However, GenDPR's distributed protocol remains faster than the centralized baseline approach in all settings.

Table 4: Comparison of the selected SNPs after each phase of the privacy-protecting verification.

| # of genomes / SNPs | # of retained SNPs | | |
|---|---|---|---|
| | Centralized | GenDPR | Naïve distributed |
| 7,430 / 1,000 | MAF 731 / LD 44 / LR 44 | MAF 731 / LD 44 / LR 44 | MAF 731 / **LD 29 / LR 29** |
| 7,430 / 2,500 | MAF 1,559 / LD 107 / LR 107 | MAF 1,559 / LD 107 / LR 107 | MAF 1,559 / **LD 66 / LR 12** |
| 7,430 / 5,000 | MAF 2,666 / LD 208 / LR 208 | MAF 2,666 / LD 208 / LR 208 | MAF 2,666 / **LD 127 / LR 29** |
| 7,430 / 10,000 | MAF 4,584 / LD 375 / LR 375 | MAF 4,584 / LD 375 / LR 375 | MAF 4,584 / **LD 240 / LR 240** |
| 14,860 / 1,000 | MAF 303 / LD 25 / LR 25 | MAF 303 / LD 25 / LR 25 | MAF 303 / **LD 11 / LR 11** |
| 14,860 / 2,500 | MAF 1,032 / LD 50 / LR 50 | MAF 1,032 / LD 50 / LR 50 | MAF 1,032 / **LD 22 / LR 22** |
| 14,860 / 5,000 | MAF 2,021 / LD 105 / LR 105 | MAF 2,021 / LD 105 / LR 105 | MAF 2,021 / **LD 44 / LR 44** |
| 14,860 / 10,000 | MAF 3,767 / LD 187 / LR 187 | MAF 3,767 / LD 187 / LR 187 | MAF 3,767 / **LD 80 / LR 80** |

The LR-test analysis phase takes the longest amount of time due to the fact that besides operating on larger data structures (2D matrix instead of 1D vectors as in previous phases), GenDPR uses an empirical approach when selecting the safe SNP-subset among the available SNPs similarly to SecureGenome [41]. Such an approach requires several iterations over several sets of SNPs. In general, GenDPR only imposes slightly longer running time due to the extra coordination and aggregation tasks performed by the leader, but the presence of more members in the federations can actually improve GenDPR's running time. GenDPR's running time depends on the distribution of the genome data being assessed in the analysis. For instance, for some populations more or fewer SNPs are removed at each phase. In particular, a higher number of retained SNPs through the phases means increased running time since statistics need to be computed over a larger space. Yet, doing so means also more from the original interest set of SNPs can be published in terms of the results of the performed analysis.

### 7.3 Correctness and Effectiveness

To assert the correctness and effectiveness of our approach, we compared the SNP positions selected as safe by GenDPR, the centralized baseline and a naïve distributed protocol. In the naïve approach, each GDO computes the LD and LR-test independently (relying only on their local dataset) and shares an encrypted vector of selected SNP indexes, of which the leader computes an intersection and outputs as safe only mutually chosen SNPs. On the other hand, in GenDPR the LD and LR-test analyses are run locally by each GDO leveraging the allele frequency vectors shared by the leader. Recall, correct LD verification needs pooling pairwise allele statistics over all individuals and the LR-test requires pooling all genomes to produce the LR-matrix used in the test.

Table 4 presents the number of SNPs retained as safe after each phase of GenDPR's privacy-protection obtained considering 7,430 or 14,860 case genomes and several number of SNPs. First, we noticed that changing the number of GDOs in the federation does not affect the outcome of the verification, as expected. In addition, we can see that GenDPR imitates the behavior of the centralized baseline over all verification phases, which shows that GenDPR is correct and does not suffer from perturbation throughout its execution.

Moreover, if intermediate data is not aggregated and considered correctly, this can lead to incorrect SNP selections. Indeed, we detected that even though such a scheme is able to retain the same SNPs during the MAF evaluation, it is not able to correctly perform the LD and LR-test analyses since the latter verifications need to consider the global genome distribution to correctly identify safe SNPs, which is not enforced with a naïve aggregation. We can identify this behavior in the bold lines of Table 4, where the naïve protocol inappropriately identified a smaller and disjoint set of SNPs. The release of such SNPs would allow membership inference of participants in the study. On the other hand, the adjustments we render in GenDPR thwart such issues, i.e., GenDPR selects the same set of SNPs as the centralized baseline.

Table 5: Collusion-tolerant GenDPR results considering 10,000 SNPs and 14,860 genomes.

| Settings | # safe released SNPs **with** collusion-tolerance | # vulnerable SNPs **without** collusion-tolerance | Running time (ms) |
|---|---|---|---|
| $G = 3, f = 1$ | 141 (75.4%) | 46 (24.6%) | 123,338.5 |
| $G = 3, f = 2$ | 143 (76.5%) | 44 (23.5%) | 76,362.5 |
| $G = 3, f = \{1, 2\}$ | 138 (73.8%) | 49 (26.2%) | 158,059.5 |
| $G = 4, f = 1$ | 143 (76.5%) | 44 (23.5%) | 159,293.2 |
| $G = 4, f = 2$ | 139 (74.3%) | 48 (25.7%) | 156,569.9 |
| $G = 4, f = 3$ | 145 (77.5%) | 44 (22.5%) | 80,681.4 |
| $G = 4, f = \{1, 2, 3\}$ | 136 (72.7%) | 51 (27.3%) | 309,032.3 |
| $G = 5, f = 1$ | 144 (77.1%) | 43 (22.9%) | 215,347.1 |
| $G = 5, f = 2$ | 135 (72.1%) | 52 (27.9%) | 255,071.8 |
| $G = 5, f = 3$ | 137 (73.3%) | 50 (26.7%) | 181,159 |
| $G = 5, f = 4$ | 148 (79.1%) | 39 (20.9%) | 79,300.4 |
| $G = 5, f = \{1, 2, 3, 4\}$ | 134 (71.7%) | 53 (28.3%) | 605,281.8 |

## 7.4 Collusion-Tolerant GenDPR

Table 5 evaluates the impact of collusion-tolerance on GenDPR in terms of privacy (detecting SNPs that would become vulnerable given the presence of colluders) and performance (running time and release coverage). Between 20.9% and 28.3% of the SNPs are vulnerable when members collude. GenDPR can seclude these vulnerable SNPs and refrains from releasing statistics over them. Thus, we see an expected impact on the number of SNPs being released proportional to the number of vulnerable SNPs. Nonetheless, collusion-tolerant GenDPR still releases between 71.7% and 79.1% of the data compared to the experiments we performed without collusion ($f = 0$) (see Table 4).

Overall, there is an increase in running time of collusion-tolerant GenDPR due to the extra verifications conducted over GDOs' isolated data. Comparing the most conservative setting of GenDPR where all possible combination of colluders are considered, i.e., $f = \{1, ..., G − 1\}$ with the $f = 0$ case, we noticed longer running times. For instance, the $G = 5, f = \{1, 2, 3, 4\}$ setting took up to 605 seconds compared to $f = 0$ with 44 seconds. Still we believe this increase in running time is a reasonable trade-off to accept for higher-levels of privacy.

Table 5 shows that shorter running times are achieved in the $f = G − 1$ setting compared to smaller $f$ values. In this scenario the additional rounds of verifications only need to be performed considering each GDO dataset individually, and therefore over fewer combination of genomes. We note that the number of safe SNPs depends on the distribution of the genome data, which impacts the identification power of participants during the LR-test evaluation. Therefore, there is no direct correlation between the number of genomes/SNPs and the number of safe SNPs. Overall, we noticed a similar behavior over the experiments that consider 1,000 SNPs.

## 8 CONCLUSION

In this work, we presented GenDPR, a distributed protocol that demonstrates that ensuring private releases of federated GWAS can be accurately and efficiently performed in a distributed fashion, in contrast to existing solutions that require centralizing genomes. Our distributed architecture removes the need for genomic data outsourcing across boundaries and so contributes to adhering to 21$^{st}$-century data privacy guidelines. GenDPR is a scalable solution compatible with the current requirements of security and offers end-to-end privacy for federated GWAS. GenDPR enables not only donors' privacy, by impeding genomic privacy attacks, but also protects institutions' intermediate data under the threat of colluding adversaries. In future work, we plan to extend GenDPR to cope with side-channel attacks against TEEs by designing an oblivious version of the protocol.

## ACKNOWLEDGMENTS

This work was supported by the European Union under the H2020 Programme Grant Agreement No. 830929 (CyberSec4Europe).